\documentclass[10pt,conference]{IEEEtran}
\IEEEoverridecommandlockouts
\usepackage{graphicx} % Required for inserting images
\usepackage{enumitem}
\usepackage{balance}
\usepackage{tcolorbox}

\usepackage{amsmath}
\usepackage{booktabs} 

\def\BibTeX{{\rm B\kern-.05em{\sc i\kern-.025em b}\kern-.08em
    T\kern-.1667em\lower.7ex\hbox{E}\kern-.125emX}}

\begin{document}

\title{Envisioning Future Interactive Web Development: Editing Webpage with Natural Language}

\author{  
    \IEEEauthorblockN{Truong Hai Dang}  
    \IEEEauthorblockA{\textit{Singapore Management University} \\
        Singapore \\
        hdtruong.2022@scis.smu.edu.sg}  
\and  
    \IEEEauthorblockN{Jingyu Xiao}  
    \IEEEauthorblockA{\textit{The Chinese University of Hong Kong} \\
        Hong Kong SAR\\
        jyxiao@link.cuhk.edu.hk}  
\and  
    \IEEEauthorblockN{Yintong Huo*}  
    \IEEEauthorblockA{\textit{Singapore Management University} \\
        Singapore \\
        ythuo@smu.edu.sg}  
        \thanks{$^{\ast}$Yintong Huo is the corresponding author.}
}

% \author{\IEEEauthorblockN{Dang $^{1}$, Jingyu Xiao$^{2}$, Yintong Huo$^{1\ast}$}

% \IEEEauthorblockA{$^1$Singapore Management University, Singapore$^2$ The Chinese University of Hong Kong, Hong Kong}

% \thanks{$^{\ast}$ Yintong Huo is the corresponding author.}

\maketitle
\begin{abstract}
The evolution of web applications relies on iterative code modifications, a process that is traditionally manual and time-consuming. While Large Language Models (LLMs) can generate UI code, their ability to edit existing code from new design requirements (e.g., "center the logo") remains a challenge. This is largely due to the absence of large-scale, high-quality tuning data to align model performance with human expectations.
In this paper, we introduce a novel, automated data generation pipeline that uses LLMs to synthesize a high-quality fine-tuning dataset for web editing, named Instruct4Edit. Our approach generates diverse instructions, applies the corresponding code modifications, and performs visual verification to ensure correctness.  By fine-tuning models on Instruct4Edit, we demonstrate consistent improvement in translating human intent into precise, structurally coherent, and visually accurate code changes. 
This work provides a scalable and transparent foundation for natural language–based web editing, demonstrating that fine-tuning smaller open-source models can achieve competitive performance with proprietary systems.
We release all data, code implementations, and model checkpoints for reproduction\footnote{https://github.com/dangtruong01/Instruct4Edit}.
\end{abstract}

% \begin{document}

% \maketitle

\section{Introduction}
Transforming webpage designs into UI code is a crucial but time-consuming stage in web development~\cite{Moran2018AutomatedRO}. Recent advances in Multimodal Large Language Models (MLLMs) have shown impressive capabilities in generating code from visual inputs~\cite{gui2025webcode2m, xiao2025designbench, tang2025slidecoder, yang2024swe, liu2025logomotion, lu2025misty, li2024mmcode, Agile}, opening new possibilities for automating design-to-code conversion~\cite{wan2025divide, xiao2024interaction2code, xiao2025efficientuicoder, wan2025automatically}.

However, editing existing code, rather than creating it from scratch, is central to how the web application evolves. Developers rarely build entire interfaces in a single pass, instead, they iteratively refine existing codebases in response to new design requirements such as ``make the layout more minimalist'', ``increase spacing here'', or ``center the logo''.
These modifications are time-consuming for developers and demand deep understanding of the web structure and content. To accelerate this process for software evolution, this paper envisions a future web application development paradigm: \textit{interactive web UI editing through natural language} (i.e., web editing in this paper).
% While common in practice, these tasks remain callenging for MLLMs, which struggle to interpret abstract instructions and apply coherent updates to real-world code.

\begin{figure}[tbp]
    \centering
    \includegraphics[width=1\linewidth]{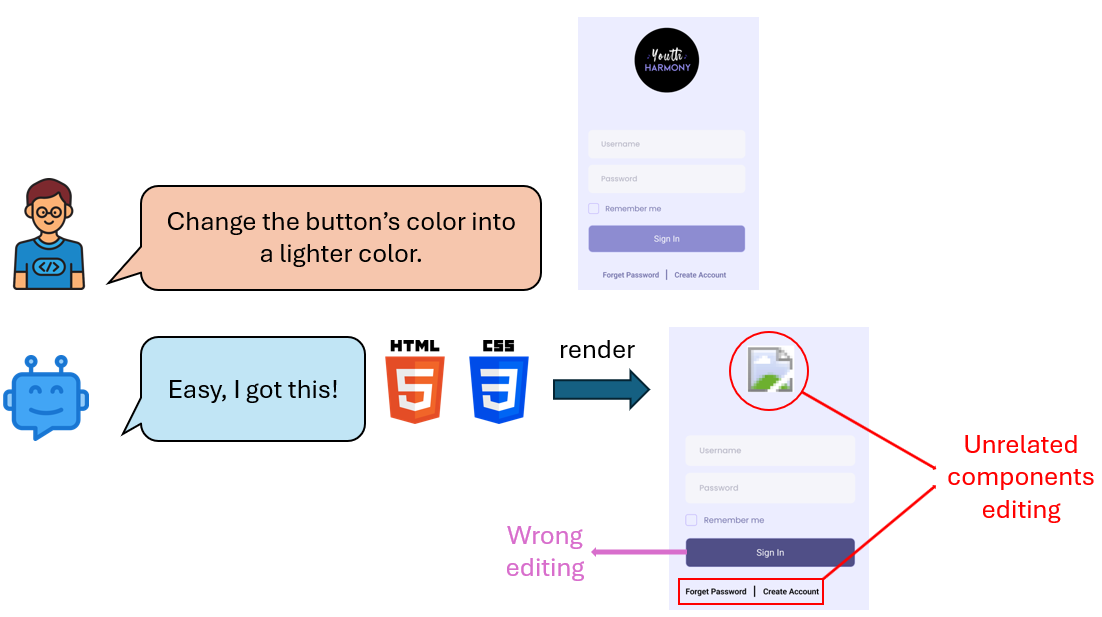}  % adjust width and filename
    \caption{Example of a Failed HTML Edit Based on a Design Instruction}
    \label{fig:ui-edit}
\end{figure}

While LLMs have recently excelled at generating UI code, their capacity to edit real-world code based on instructions remains unsatisfied (shown in Fig.~\ref{fig:ui-edit}). 
This task is challenging because it requires LLMs to fully understand HTML, reason about GUI layouts, and comprehend human intent to apply changes. Specifically, we identify three key challenges in automatic web editing:

% few systems can reliably handle instruction-driven editing.
%Current models often require complex prompts or produce differences instead of full HTML rewrites, and they frequently struggle to apply stylistic or layout changes in a visually grounded manner. 
% This limitation stems from several challenges: 
\begin{itemize}[leftmargin=*]
    \item \textbf{Abstract design instructions} such as ``make the layout more minimalist'' are inherently vague and require grounding in visual semantics.
    \item Models must generate \textbf{coherent HTML} to apply changes while keeping the rest of the webpage stable, which demands structural consistency and correctness.
    \item The output must exhibit \textbf{visual alignment}, ensuring that the rendered webpage reflects the user's intended modification. 
\end{itemize}

A core reason for the lack of reliable models is the absence of large-scale, high-fidelity datasets that pair natural language instructions with corresponding HTML/CSS modifications in an end-to-end fashion. Such datasets enable LLMs to learn UI domain knowledge and align with user preferences. However, creating these datasets manually is impractical due to the extensive effort from designers, developers, and annotators to craft instructions, implement edits, and verify correctness. 

To bridge this gap, we propose a novel paradigm for automatically synthesizing instruction-tuning datasets. Our pipeline leverages large language models across the full supervision flow—first by eliciting latent design-editing knowledge to generate diverse, human-like instructions, then by producing corresponding HTML code modifications, and finally by conducting automated visual cross-checking to validate whether the intended edits have been correctly applied. This self-contained loop allows LLMs to act both as synthetic supervisors and evaluators, transforming implicit UI-editing knowledge into explicit, verifiable training data. The resulting dataset, \textbf{Instruct4Edit}, consists of high-quality (instruction, original HTML, modified HTML) samples that are semantically aligned and visually faithful, offering a scalable foundation for fine-tuning code-editing agents.

Using Instruct4Edit, we fine-tune LLMs to perform single-shot HTML rewrites grounded in natural language commands. These models show improved performance in both structure preservation and visual correctness, demonstrating that high-quality synthetic data can effectively translate human intent into concrete code changes. The goal is not to outperform frontier commercial systems, but to demonstrate that smaller open-source models can achieve competitive performance through targeted fine-tuning. Our contributions are as follows:
\begin{itemize}[leftmargin=*]
    \item We envision a natural language-enabled web editing paradigm for future interactive software development.
    % We propose a scalable, LLM-driven pipeline for synthesizing high-quality webpage edit datasets;
    \item We develop a transparent, fully-automated data synthesis pipeline that trains LLMs to follow user instructions for web editing.
    % We release Instruct4Edit, a dataset focused on web UI instruction editing;
    \item We evaluate our pipeline on real-world editing jobs, showing approximately a notable 10\% improvement for existing models.
\end{itemize}

\section{Related work} %% motivated example
\textbf{Dataset Synthesis for Instruction Tuning.}
Instruction tuning of LLMs relies on large-scale datasets where instructions are aligned with desired outputs. 
% Manually creating such datasets for HTML and CSS is costly, involving data gathering, code rewriting, and quality checks. 
Existing datasets like WebSight~\cite{laurenccon2024unlocking} synthesize screenshot-to-code generation without human-provided instructions. 
While SelfCodeAlign~\cite{wei2024selfcodealign} offers synthetic validation pipelines, it targets functional code rather than UI semantics. 
Our approach is novel in using an LLM-driven pipeline to synthesize a high-quality dataset for enhancing web UI editing.
% verified (instruction, original HTML, modified HTML) triplets. This enables Instruct4Edit, a high-quality instruction tuning dataset for full-page web UI edits for LLM models.

\textbf{UI Code Intelligence.}
Most existing research~\cite{yun2024web2code, si2025design2code, gui2025UICopilot, moran2018machine, nguyen2015reverse, Chen2018FromUI, gui2024vision2ui} focuses on automated UI code generation from visual inputs like screenshots.
For instance, DeclarUI~\cite{zhou2024bridging}, LayoutCoder~\cite{wu2024mllm} and DCGen~\cite{wan2024automatically} use visual segmentation to map layouts to HTML code for one-shot generation from scratch.Other works like Interaction2Code~\cite{xiao2024interaction2code} generate the interactive applications in iterative refinement with a focus on functionality testing and DesignRepair~\cite{designrepair} and Nighthawk~\cite{liu2022nighthawk, liu2020owl} focus on the UI issues detection and repairing. Previous research does not address the continuous instruction-following needs of ongoing web editing.

\section{Problem Formulation}

% This work addresses the design edit task, where the input is a natural language instruction $I$ describing a visual design change, and an existing HTML code file $C$ that renders a web page

% a visual design of a webpage and , and the goal is to generate HTML+CSS code that accurately reproduces it. Let $I_0$ be the design image of a webpage, and $C_0$ be the ideal HTML+CSS code. Given $I_0$, an MLLM $M$ generates code $C_g = M(I_0)$, and the rendered output $I_g$ of $C_g$ should closely match $I_0$ in both structure and appearance.

This work addresses the design editing task, where the inputs are a natural language instruction $I$ describing a visual design change, and an existing HTML code file $C_0$ that renders a webpage $W_0$, the goal is to produce a fully modified HTML code file that accurately reflects the intended edit.  Given $I$ and $C_0$, a LLM $M$ generates code $C_g = M(I, C_0)$ to render the modified webpage $W_g$. 
% This task is single-shot and document-level, meaning the model must generate a complete HTML rewrite without relying on diffs, patches, or step-by-step interaction.
The rendered webpage $W_g$ must preserve instruction-unrelated portions of the original HTML while modifying only the relevant components, ensuring \textit{semantic fidelity}, \textit{visual correctness}, and \textit{structural integrity}.

\section{Methodology}
We present a two-stage framework for enabling instruction-grounded HTML rewriting: (1) a scalable, automated pipeline for synthesizing high-quality design-edit instructions paired with corresponding HTML modifications, and (2) fine-tuning large language models on the resulting dataset. The core insight lies in leveraging LLMs throughout the entire supervision loop — not only to simulate human-like design modification instructions and generate coherent code edits, but also to verify visual fidelity through cross-modal evaluation. This section outlines the full data synthesis pipeline and the subsequent model adaptation strategy.
\subsection{Dataset Generation}
\subsubsection{Pipeline}
\begin{figure*}[tbp]
    \centering
    \includegraphics[width=0.8\linewidth]{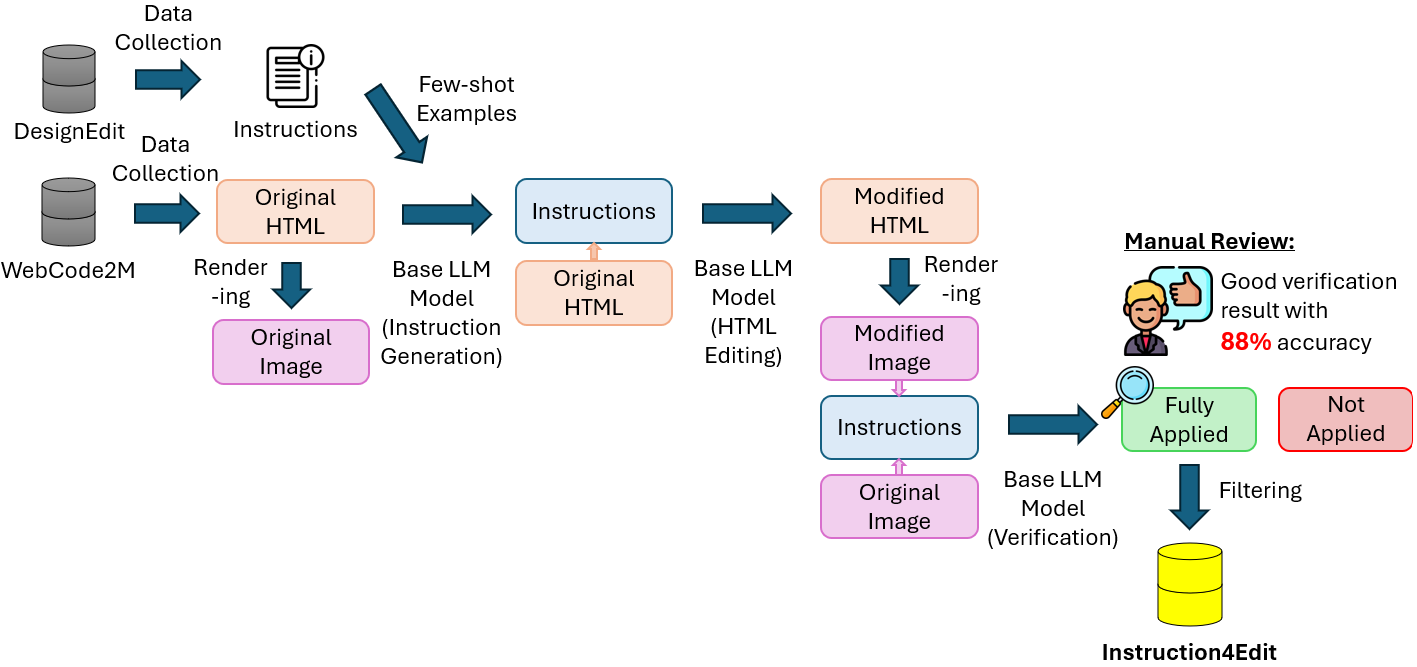}  % adjust width and filename
    \caption{End-to-end Pipeline to synthesize dataset with LLMs}
    \label{fig:pipeline}
\end{figure*}

We build \textbf{Instruct4Edit}, a clean and diverse dataset of instructional HTML edits, without involving human annotators. To avoid the high cost and scalability issues of manual data creation, we design a fully automated data generation pipeline that consists of three LLM-based components (shown in Fig~\ref{fig:pipeline}): instruction generator, editor, and verifier. Technically, all three components can be implemented using the same underlying LLM.

\textbf{Data Collection.}
We build our pipeline upon a set of real-world UIs. Specifically, we choose \textit{WebCode2M~\cite{gui2025webcode2m}} dataset, which provides high-quality and real HTML/CSS code paired with rendered webpage screenshots. From this dataset, we randomly sample 500 examples to serve as seeds in this step.

\textbf{Instruction Generation.}  
Then, we generate natural language edit instructions that capture human design intent, simulating real-world modification requirements such as (e.g., ``add spacing between sections''). To achieve this, we follow existing research \cite{wei2024selfcodealign} by supplying the LLM with a few carefully selected examples drawn from the \textit{DesignEdit} dataset \cite{xiao2025designbench} that encourage diverse instruction generation. These examples provide diverse, human-authored design-edit pairs that help the model learn how 
to phrase visually grounded modification requests in natural language. Specifically, we provide the generator with original HTML and examples from
% prompt \textit{Gemini-2.5-Pro} with the original HTML, guided by curated examples from 
prior real-world UI edits to avoid any mention of code or syntax. Each sample produces five diverse, human-like instructions.

\textbf{HTML Editing.}  
The next step is to apply each instruction to the original HTML in a coherent and complete way. Instead of generating differences, we request the editor to return a fully rewritten HTML document. This ensures the output remains renderable and self-contained.

\textbf{Rendering \& Verification.}  
To verify the quality of the edits, we render both the original and modified HTML files in a fixed viewport setting for consistency. Then, these image pairs, along with the original instruction, are passed to the verifier that acts as a strict visual reviewer. The verifier component is implemented using an LLM configured as a visual-text reasoning agent. It then conducts ``cross-modal verification'' by comparing visual changes with original instructions, and classifies the edit as fully applied, or not applied. Only samples where the verification step confirms visual alignment are included in the final dataset. Each retained example includes: a human-like instruction $I$, an original HTML document $C_0$, and a modified HTML document $C_M$.

\subsubsection{Manual Quality Review}
% \textbf{Manual Review.}  
While LLM verification accelerates filtering, it may introduce noise. So we conduct a manual inspection to validate the filtering quality. Two independent human reviewers cross-checked a random subset of 50 accepted samples by checking if the edit instruction was applied in the modified HTML content. This step confirms that the automatic verification yields \textbf{88\%} agreement with human checking, indicating the high quality of our dataset.
Inter-annotator reliability between the two reviewers was measured using Cohen's Kappa, 
yielding $\kappa \approx 0.84$, which corresponds to ``almost perfect agreement'' according to the benchmark interpretation by Landis \& Koch ~\cite{landis1977measurement}. 
Disagreements were resolved through consensus discussion.

\subsubsection{Dataset Statistics}
% \textbf{Filtering and Output.} 
% Only samples where the verification step confirms visual alignment are included. Each retained example includes: a human-like instruction $I$, an original HTML document $C_0$, and a modified HTML document $C_M$.

We begin with 500 seed HTML files sampled from WebCode2M. For each, we generate 5 instructions, resulting in 2,500 instruction–HTML pairs. These pairs are processed through our editing and rendering pipeline, yielding 2,500 modified HTML candidates. After automated verification, 1,150 samples are accepted with an acceptance rate of \textbf{46\%}. Most rejections are due to partial instruction alignment, visual rendering mismatches, or edits targeting hidden components (e.g., CSS-only visibility). We found these cases to be semantically ambiguous, suggesting opportunities for refining verifier precision in future iterations.

\subsubsection{Dataset Implementation Details} To enhance reproducibility, we provide additional implementation details of \textit{Instruct4Edit}.

\textbf{Source corpus.} We used \textit{WebCode2M~\cite{gui2025webcode2m}}, which contains approximately 2.56 million webpage code pairs with realistic layouts and styling.

\textbf{Sampling.} We randomly sampled 500 HTML/CSS examples from \textit{WebCode2M} to serve as base templates for generating edits.

\textbf{Instruction generation.} For each base sample, we generated five natural-language editing instructions using the Gemini-2.5-Pro model, resulting in 2,500 instruction--HTML pairs. Prompts were explicitly constrained to avoid mentioning technical identifiers (e.g., class or ID names) to promote natural, human-like phrasing. Example instruction types include:
\begin{itemize}
    \item \textit{Layout edits:} ``Make the layout more minimalist.''
    \item \textit{Spacing edits:} ``Increase padding between sections.''
    \item \textit{Styling edits:} ``Change the header font to something bold and modern.''
    \item \textit{Color edits:} ``Use a softer background color.''
\end{itemize}

\textbf{Token statistics.} The combined length of each instruction--HTML input sequence averaged approximately 2{,}800 tokens (median $\approx$ 2{,}500; maximum $\approx$ 5{,}500), motivating our choice of Qwen2.5-7B due to its extended context window (128{,}000 tokens).

\textbf{Training setup.} Fine-tuning was performed on a single NVIDIA A100 (80~GB) using LoRA adapters, with a batch size of~8 and peak memory usage around~60~GB. This setup ensures efficient training without gradient checkpointing or model parallelism.

Our pipeline ensures that all samples in Instruct4Edit are verifiable and consistent with human instructions, thus reducing noise and enhancing the robustness of the instruction-tuning process.

\subsection{Tuning Models with Efficiency}

% instruction-following variants of the \textbf{Qwen2.5-7B family~\cite{qwen2.5-vl}} based on Instruct4Edit dataset, which contains validated (instruction, original HTML, modified HTML) triplets. 
Although pre-trained LLMs have strong generative abilities, they often struggle with precise, context-aware edits, especially for ambiguous or visually related instructions. Tuning on instruction-aligned examples helps models better align their output with user intent. To enable full-page HTML rewriting from high-level design instructions, we fine-tune LLMs using (instruction, original HTML, modified HTML) triplets from the Instruct4Edit dataset.

% We focus primarily on the text-only model variant: \textbf{Qwen2.5-7B-Instruct}, which offers a favorable trade-off between performance and efficiency in handling long-form inputs. 
For efficiency, we apply LoRA adapters~\cite{hu2022lora} during tuning - a parameter-efficient technique that injects low-rank trainable weights into frozen pre-trained models- framing each task as single-shot text generation. The model input includes the original HTML and a natural language instruction, while the output is the fully modified HTML file.

\section{Experiment}
To explore the future of interactive web development via Instruct4Edit, we evaluate our approach from two perspectives: (RQ1) How effectively does Instruct4Edit support interactive edits? (RQ2) How many edit requirements are successfully applied, verified by humans? 
Our goal is to assess whether fine-tuning on Instruct4Edit improves instruction-grounded editing performance for any LLM, using Qwen2.5-7B as a representative open-source model.

\subsection{Experiment Settings}

\subsubsection{Evaluation Dataset}
For evaluation, we additionally sampled webpages from WebCode2M with the generated diverse instructions, ensuring these samples were excluded from the training data. Each edit request was manually implemented by an expert, and then verified by a second verifier to confirm that the changes were correctly applied to HTML.

\subsubsection{Baselines}
We choose the Qwen2.5 series~\cite{bai2025qwen2} instead of alternatives like LLaMA2-7B~\cite{touvron2023llama} because it supports a larger maximum token limit.
This is essential for our use case, where entire HTML documents serve as both input and output. 
Models with limited context windows, such as LLaMA-series, would require truncation, breaking the structural integrity.
Specifically, we tune Qwen2.5-7B on the Instruct4Edit dataset, serving as our main approach, denoted as Qwen2.5-7B-Instruct.

Additionally, we include two open-sourced models (tuning-free) and two large commercial LLMs (inference-only) as baselines for comparison, to further underscore the contribution of Instruct4Edit: 
(1) Qwen2.5-7B: receives only the instruction and original HTML as input.
(2) Qwen2.5-7B-VL (Multimodal): additionally incorporates a rendered screenshot of the original HTML.
(3) Gemini-2.5-Pro: a proprietary multimodal model accessed via API, used here in a zero-shot setting.  
(4) GPT-4o-mini: a commercial instruction-tuned model with vision capabilities, also evaluated with zero-shot. All prompts are included in the replication package.

% \begin{itemize}
%     \item \textbf{Qwen2.5-7B (Text-only):} receives only the instruction and original HTML as input.
%     \item \textbf{Qwen2.5-7B-VL (Multimodal):} additionally incorporates a rendered screenshot of the original HTML.
% \end{itemize}  

% As external references, we include inference-only baselines using two commercial LLMs — \textbf{Gemini-2.5-Pro\cite{comanici2025gemini}} and \textbf{GPT-4o-mini\cite{gpt4o-mini}} — to compare against zero-shot general-purpose models with strong instruction-following capabilities.

\subsubsection{Metrics}
Evaluation is conducted using both automatic metrics (RQ1) and human judgment (RQ2) to assess results. Specifically, we adopt two visual metrics following previous work:
\begin{itemize}[leftmargin=*]
    \item \textbf{Structural Similarity Index Measure (SSIM)~\cite{wang2004image}:} Captures perceptual differences between the original and modified renderings, focusing on layout consistency and low-level structural preservation. Given two grayscale images \( I_1 \) and \( I_2 \), SSIM is defined as:

    \[
    \text{SSIM}(I_1, I_2) = \frac{(2\mu_1\mu_2 + C_1)(2\sigma_{12} + C_2)}{(\mu_1^2 + \mu_2^2 + C_1)(\sigma_1^2 + \sigma_2^2 + C_2)}
    \]
    
    where \( \mu \), \( \sigma^2 \), and \( \sigma_{12} \) denote local means, variances, and covariances of the image patches, and \( C_1, C_2 \) are constants to stabilize the division. SSIM scores range from 0 to 1, with higher values indicating stronger structural similarity.

    \item \textbf{CLIP-Based Semantic Similarity~\cite{radford2021learning}:} Evaluates high-level perceptual and semantic consistency using the CLIP ViT-B/32 model \cite{radford2021learning}. Let \( f(I) \) be the CLIP embedding of image \( I \), then cosine similarity is calculated as:

\[
\text{CLIP}(I_1, I_2) = \frac{f(I_1) \cdot f(I_2)}{\|f(I_1)\| \|f(I_2)\|}
\]

We normalize this to the \([0,1]\) range by computing \( (\text{cosine\_sim} + 1) / 2 \). 
\end{itemize}

\subsection{Implementation} 
We implement our tuning pipeline with PyTorch and HuggingFace’s transformers~\cite{wolf2020transformers}, utilizing peft library to support LoRA. All experiments are conducted on a single NVIDIA A100 GPU with 80GB memory. Each model variant is trained for 3 epochs with a batch size of 8, a learning rate of 2e-5, and a maximum sequence length of 8192 tokens to handle full HTML documents without truncation. Average inference time per edit is 1~2 minutes, enabling practical deployment. We use Gemini-2.5-Pro as the base LLM model for dataset generation.

\subsection{Experiment results}
\subsubsection{How effectively does Instruct4Edit support interactive edits?}
We report quantitative results with SSIM (for structural layout similarity) and CLIP (for semantic visual similarity) between the original and modified HTML renderings. Table~\ref{tab:quantitative-results} summarizes the average scores of our models across the evaluation, with Qwen2.5-7B-Instruct achieving the \textbf{highest visual similarity in both structural and semantic metrics}.

\begin{table}[t]
\centering
\caption{Quantitative Evaluation of Editing Results}
\label{tab:quantitative-results}
\begin{tabular}{lcc}
\toprule
\textbf{Model} & \textbf{SSIM} & \textbf{CLIP} \\
\midrule
GPT-4o-mini          & 0.896 & 0.987 \\
Gemini-2.5-Pro       & 0.883 & 0.979 \\
Qwen2.5-7B-VL        & 0.764 & 0.960 \\
Qwen2.5-7B-Base      & 0.796 & 0.975 \\
Qwen2.5-7B-Instruct (Ours)  & \textbf{0.952} & \textbf{0.993} \\
\bottomrule
\end{tabular}
\end{table}

% The instruction-tuned Qwen2.5-7B-Instruct achieves the highest visual similarity in both structure and semantic similarity among all baselines. 

% This result demonstrates the effectiveness of our instruction-aligned dataset and finetuning strategy in enabling web editing.

The gap between the tuned and basic (w/o tuned) Qwen2.5-7B model (0.952 vs. 0.796 SSIM, 0.993 vs. 0.975 CLIP) highlights the significance of Instruct4Edit. The model learns to apply semantically aligned edits that preserve the original layout unless instructed otherwise.

Surprisingly, the Qwen2.5-7B-VL model underperforms its textual counterpart, suggesting that \textbf{raw visual input might provide limited benefit for this task~\cite{xiao2025designbench}}. Since HTML already encodes structure and semantics, adding screenshots may introduce redundancy.
Therefore, we regard the vision-free approach as superior, both in effectiveness and efficiency, as it reduces memory usage and accelerates inference, making it better suited for practical real-world deployment in the future.

Despite their large model sizes, both GPT-4o-mini and Gemini underperform compared to our fine-tuned model. This highlights that, even with strong base models, \textbf{domain-specific fine-tuning can yield substantial improvements for structured generation tasks} such as UI code editing.

\subsubsection{How many edit requirements are successfully applied, verified by humans?}
To complement the automated evaluation, we conducted a manual evaluation on the same set of 50 design-edit samples, judging whether the modified HTML satisfied the instruction intent. Each output was independently labeled as a \textit{pass} (instruction correctly applied) or \textit{fail} (instruction not fully reflected or misapplied).

\begin{table}[t]
\centering
\caption{Human evaluation results across models.}
\label{tab:human-eval}
\begin{tabular}{lccc}
\toprule
\textbf{Model} & \textbf{Passes} & \textbf{Fails} & \textbf{Passing Rate (\%)} \\
\midrule
GPT-4o-mini & 29 & 21 & 58 \\
Gemini-2.5-pro & 26 & 24 & 52 \\
Qwen2.5-7B-VL & 18 & 32 & 36 \\
Qwen2.5-7B-Base & 24 & 26 & 48 \\
Qwen2.5-7B-Instruct (Ours) & 28 & 22 & 56 \\
\bottomrule
\end{tabular}
\end{table}

As shown in Table~\ref{tab:human-eval}, fine-tuning the base Qwen2.5-7B model with Instruct4Edit significantly \textbf{improves instruction-following ability}, increasing the pass rate from 48\% to 56\%. This confirms that even a relatively small, high-quality dataset like Instruct4Edit can enhance the model's precision in UI-specific code editing.

The vision-enabled variant (Qwen2.5-7B-VL) underperforms its text-only version, with the lowest pass rate (36\%) among all evaluated models. This aligns with our earlier observation that visual inputs may not introduce meaningful grounding for code-focused tasks, and could even distract generation.

% Among commercial models, GPT-4o-mini leads with 58\% pass rate, marginally outperforming our fine-tuned Qwen model. This highlights the strong general instruction-following ability of large-scale foundation models, though our approach remains competitive given its significantly smaller footprint and open accessibility.

While GPT-4o-mini achieves a slightly higher pass rate (58\%) than our fine-tuned Qwen model (56\%), its advantage is marginal. Our approach achieves \textbf{competitive results with a significantly smaller model size and open-source accessibility}, making it a more practical and customizable choice for real-world deployment in UI editing.

% \begin{tcolorbox}
% [
%     colback=grey!5!white,%,
%     colframe=grey!75!black,%,
%     fonttitle=\bfseries,
%     sharp corners
% ]
% \small
% \textbf{Summary to RQs:} 
% With our high-quality tuning dataset, the models surpass baselines by a wide margin of [xx\%] (SSIM/CLIP) and [xx\%] (passing rate), demonstrating the effectiveness of Instruct4Edit on the real-world web editing problem.
% \end{tcolorbox}
\begin{figure}[t]
\centering
\includegraphics[width=\linewidth]{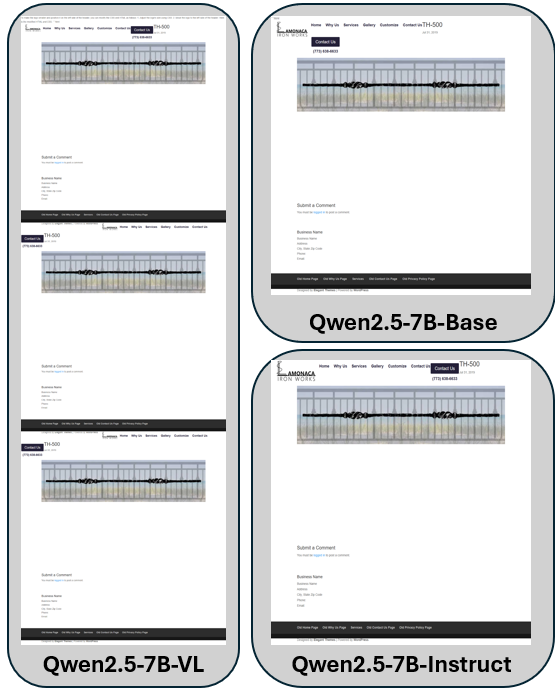}
\caption{Design edit outputs across model variants}
\label{fig:case-study}
\end{figure}

\section{Case study}
To illustrate the practical performance of evaluated models, we present a case study where three baselines are tasked with the same instruction in Fig~\ref{fig:case-study}: \textit{Make the logo smaller and position it on the left side of the header}.
As shown in Figure~\ref{fig:case-study}, the base Qwen2.5-7B shifts unrelated components and misaligns the overall layout, while the Qwen2.5-7B-VL repeats the entire page multiple times vertically, failing to maintain the current design. In contrast, our fine-tuned Qwen2.5-7B-Instruct correctly resizes and repositions the logo, while preserving all other UI elements. 

% This case exemplifies how targeted fine-tuning enables precise and stable instruction-following behavior in full-document HTML rewriting.

% \section{Discussion}

\section{Conclusion and Future Plan}
This paper addresses a novel and practical task in continuous web development: automated editing of web HTML code to meet new requirements.
To this end, we 
introduce Instruct4Edit, a transparent and automated dataset generation pipeline that constructs a high-quality instruction-tuning dataset to enable LLMs to better align with human expectations in web development. By fine-tuning open-source models on such dataset, we demonstrate significant improvements in layout preservation and edit satisfaction.
% that enables full-document HTML rewriting based on natural language design edits. Our work presents a scalable, fully-automated data generation pipeline that utilizes LLMs for not only instruction generation, but also code editing and visual verification - yielding clean, semantically aligned training samples without human supervision. 

% By fine-tuning open-source models on this dataset, we demonstrate strong improvements in layout preservation and instruction satisfaction. This validates the utility of high-quality, text-supervised data for HTML editing. However, the long-token nature of HTML remains a limitation, as large inputs may exceed context windows, leading to truncation or incomplete edits.

Looking forward, we plan to extend the pipeline for broader evaluation and exploit the LLM’s reasoning capabilities to enable more practical automated web editing.
Future work includes expanding our framework to support diverse front-end frameworks, such as React and Vue, allowing instruction-driven edits in component-based frontends. Additionally, we aim to incorporate relevant UI programming knowledge through retrieval-augmented generation to further improve the LLM’s reasoning process.

\section*{Acknowledgement}
The work was supported by the Singapore Ministry of Education (MOE) Academic Research Fund (AcRF) Tier 1 grant.

\balance
\bibliographystyle{IEEEtran} 
\bibliography{sample-base}

\end{document}